\documentclass[conference]{IEEEtran}
\IEEEoverridecommandlockouts
\usepackage{cite}
\usepackage{amsmath,amssymb,amsfonts}
\usepackage{algorithmic}
\usepackage{graphicx}
\usepackage{subcaption}
\usepackage{textcomp}
\usepackage{xcolor}
\usepackage{bm}

\usepackage[labelformat=simple]{subcaption}

\DeclareMathOperator{\diag}{diag}

\graphicspath{{Fig/},{fig/}}
\def\BibTeX{{\rm B\kern-.05em{\sc i\kern-.025em b}\kern-.08em
    T\kern-.1667em\lower.7ex\hbox{E}\kern-.125emX}}

\begin{document}

\newcommand{\fig}{Fig.\xspace}
\newcommand{\cm}[1]{}

\title{WiRiS: Transformer for RIS-Assisted Device-Free Sensing for Joint People Counting and Localization using Wi-Fi CSI\\
}

\author{\IEEEauthorblockN{Wei-Yu Chung, Li-Hsiang Shen, Kai-Ten Feng, Yuan-Chun Lin$^\dagger$, Shih-Cheng Lin$^\dagger$, and Sheng-Fuh Chang$^*$\\}
\IEEEauthorblockA{Department of Electronics and Electrical Engineering, National Yang Ming Chiao Tung University, Hsinchu, Taiwan\\
$^\dagger$Department of Electrical
Engineering, National Chung Cheng University, Chiayi, Taiwan\\
$^*$Department of Communications Engineering, National Chung Cheng University, Chiayi, Taiwan\\
Email: willboy0218.c@nycu.edu.tw, gp3xu4vu6.cm04g@nctu.edu.tw, ktfeng@nycu.edu.tw, \\yolenlin@alum.ccu.edu.tw, sclinee@ccu.edu.tw, ieesfc@gmail.com
}}

\maketitle

\begin{abstract}
Channel State Information (CSI) is widely adopted as a feature for indoor localization. Taking advantage of the abundant information from the CSI, people can be accurately sensed even without equipped devices. However, the positioning error increases severely in non-line-of-sight (NLoS) regions. Reconfigurable intelligent surface (RIS) has been introduced to improve signal coverage in NLoS areas, which can re-direct and enhance reflective signals with massive meta-material elements. In this paper, we have proposed a Transformer-based RIS-assisted device-free sensing for joint people counting and localization (WiRiS) system to precisely predict the number of people and their corresponding locations through configuring RIS. A series of predefined RIS beams is employed to create inputs of fingerprinting CSI features as sequence-to-sequence learning database for Transformer. We have evaluated the performance of proposed WiRiS system in both ray-tracing simulators and experiments. Both simulation and real-world experiments demonstrate that people counting accuracy exceeds 90\%, and the localization error can achieve the centimeter-level, which outperforms the existing benchmarks without employment of RIS.
\end{abstract}


\section{Introduction}

With the rise of the Internet of Things (IoT), indoor localization has become increasingly important. However, Global Positioning System (GPS) signals are unavailable in indoor environments due to obstructions and walls, and as a result, several approaches have been proposed \cite{Low_efforts, IoT, Radio_img}. The most straightforward method for indoor localization is camera-based detection, which typically provides high accuracy. However, camera-based detection is susceptible to light interference and privacy concerns. Given the high density of Wi-Fi deployment, several wireless sensing techniques have been developed for indoor positioning. Among these, the most typical indoor positioning methods are the received signal strength indicator (RSSI) \cite{ML_RSSI} and the channel state information (CSI) \cite{CSI_Loc}. RSSI-based positioning is based on the distance variation with different levels of signal attenuation. However, the main drawback of RSSI is that it fails to account for multipath effects in the complex indoor environment. In comparison, CSI has more information by leveraging the subcarriers in each orthogonal frequency division multiplexing (OFDM) symbol and is therefore more stable and has higher resolutions. In this paper, we consider CSI as a fine-grained localization feature that is more effective at representing the complex indoor environment.

However, the performance of indoor localization may be degraded in non-line-of-sight (NLoS) regions due to weak signals \cite{NLOS_Color, Improve_NLOS, my}, which can be well dealt with by the employment of reconfigurable intelligent surface (RIS). RIS is garnering increasing attention as a promising technology for the next generation of communication systems \cite{acm}. An RIS is an electromagnetic surface that consists of a large number of reflecting meta-material elements, which can reflect incident signals towards the desired direction by adjusting the phase shift of elements without requiring radio frequency (RF) chains \cite{my2}. Compared to classical antenna arrays, the RIS has the capability to customize the wireless environment and enhance signal capacity and network coverage.

In addition, the unique characteristics of RIS can also improve the performance of localization and sensing. Several works related to RIS-assisted positioning have been investigated. In \cite{WFing}, the authors proposed two solutions based on machine learning and the heuristic method to reduce the state space of the RIS to enhance the localization accuracy. In \cite{MetaRadar}, the authors adjusted the RIS configuration to reduce the similarity of received signal strength values at various positions, improving the positioning performance of multiple users. Additionally, \cite{SEMI} proposes a method to estimate the 3D position of each user equipped with an RIS, addressing both the data association problem and multipath interference. In \cite{Towards}, the authors formulated the RIS-assisted localization issue into an optimization problem that aims to increase the difference of RSSI between neighbor positions. Moreover, joint communication and positioning have also been proposed in \cite{POE}. Considering both synchronous and asynchronous schemes, authors adopt the Cramer-Rao lower bound to evaluate the performance of the positioning.
Moreover, as for sensing, in \cite{RF_sensing}, the authors propose a motion recognition system based on the RIS to enhance the recognition accuracy of motion by finding the optimal RIS configuration. Furthermore, in \cite{Meta_sensing}, the authors formulate the RF 3D sensing problem into an optimization problem, which aims to improve sensing accuracy by jointly considering the beamformer and the received signals.

Most existing works focus on a single objective, such as positioning or sensing, as demonstrated in prior research \cite{NLOS_Color, Improve_NLOS, my, WFing, MetaRadar, SEMI, Towards, RF_sensing, Meta_sensing, POE}. However, with the high channel diversity provided by RIS, it becomes possible to simultaneously estimate multiple sensing objectives. In this paper, we make several significant contributions to the field of RIS-assisted sensing and localization. Firstly, we propose a novel approach that utilizes CSI as a feature for device-free localization with the assistance of RIS, which, to the best of our knowledge, has not been attempted before. Secondly, we introduce the WiRiS scheme, which is capable of jointly predicting the number of people and their corresponding locations in non-line-of-sight regions, offering enhanced functionality compared to existing techniques. Finally, we evaluate our proposed approach through simulations and real-world experiments, which show significant improvements in people counting and localization accuracy with the assistance of RIS technology. Our work represents an important step forward in the development of RIS-assisted sensing and localization techniques and highlights the potential for future research in this area.

\section{System Architecture}

\begin{figure}
  \centering
   \includegraphics[width=0.48\textwidth]{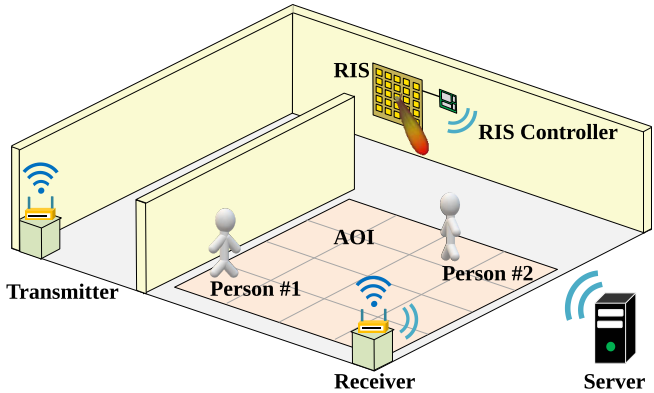}
  \caption{System architecture for RIS-assisted joint people counting and localization using commercial Wi-Fi devices.}
  \label{fig:system}
\end{figure}

\subsection{Scenario Description}
The system depicted in Fig. \ref{fig:system} models an indoor environment where a single transmitter and receiver are equipped with omnidirectional antennas. The line-of-sight (LoS) path between the transmitter and receiver is assumed to be obstructed in this scenario, which can significantly degrade the accuracy of localization over longer distances. To address this issue, our objective is to enhance the positioning performance in the area of interest (AOI) on the receiver side. To achieve this, we deploy an RIS on the wall as a reflector, which can be controlled to redirect all predefined reflected beams toward the AOI. By leveraging the RIS, we can enhance the human presence effect in the AOI. After adjusting the RIS to switch all the predefined beams, the receiver can then obtain the channel state information (CSI) data of multiple individuals standing at their respective locations.

\subsection{Channel Model}
CSI describes the propagation of multipath transmission from the transmitter to the receiver. Due to the multipath variates by people at different positions, we could receive the CSI data at the receiver as their location feature. In this paper, we consider a multi-input-multi-output (MIMO) OFDM system with $N_T$ transmit antennas, $N_R$ receive antennas, and one RIS with $M$ elements. Besides, OFDM signal consisting of $K$ subcarriers is transmitted. Let $D \in \mathbb C^{N_T \times N_R}$ denote the direct channel between the transmitter and receiver, $H_{T-RIS} \in \mathbb C^{N_T \times M}$ be the channel between the transmitter and the RIS, and $H_{RIS-R} \in \mathbb C^{N_R \times M}$ be the channel between the RIS and the receiver. The received signal $y \in \mathbb C^{N_T \times N_R}$ can be expressed as
\begin{equation} \label{received_y}
y = (H_{T-RIS}\Theta H_{RIS-R} + D)x + n,
\end{equation}
where $\Theta = \diag([\theta_1, \theta_2, \cdots, \theta_N]^T)$ represents the configuration of the RIS, and $n \in \mathbb C^{N_T \times N_R}$ is the additive white Gaussian noise (AWGN). Note that $\theta_n = \alpha_n e^{j\phi_n}$, where $\alpha_n \in [0, 1]$ and $\phi_n \in [0, 2\pi]$ denotes the amplitude and phase of each element $n$ on the RIS. Let $H_{o} \in \mathbb C^{N_T \times N_R}$ represent the effective channel between the transmitter and receiver, we could reformulate $\eqref{received_y}$ as
\begin{equation}
y = H_{o}x + n.
\end{equation}
Note that $H_{o} = |H_{o}|e^{j\angle H_{o}}$, where $|H_{o}|$ and $\angle H_{o}$ denotes the effective amplitude and phase of the overall channel. Note that in this paper, we only consider amplitude as the input feature for sensing and localization because amplitude is more stable and distinctive compared to the sensitive phases of RIS CSI.

\begin{figure}
  \centering
  \includegraphics[width=0.48\textwidth]{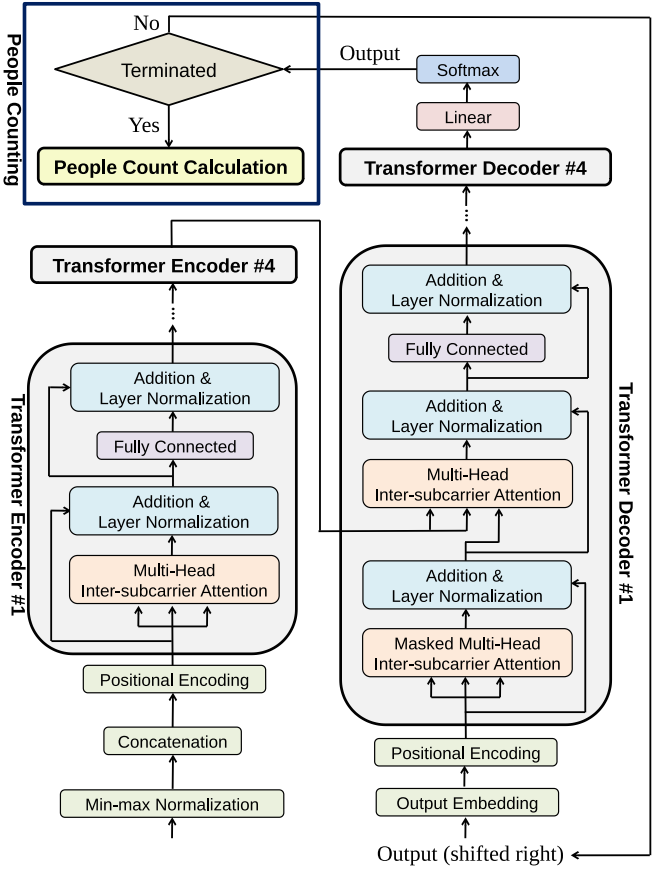}
  \caption{Schematic diagram for proposed WiRiS scheme.}
  \label{fig:block}
\end{figure}

\section{Proposed WiRiS Scheme}

Previous studies on the localization of multiple people assume that the number of people is known in advance, and the focus is solely on predicting their positions. However, in reality, it is unrealistic to have prior knowledge of the number of people present in an environment. To overcome this limitation, we propose the WiRiS system, which can jointly predict the number of people and their positions with a manner of sequence-to-sequence learning. In the next section, we describe the process of establishing the CSI fingerprinting database, along with the preprocessing techniques used for the CSI data. Finally, we present the approach we used to solve our problem using the WiRiS system.

\subsection{CSI Data Collection}
In the offline phase, we construct the CSI fingerprinting database by collecting CSI data from $S$ designated reference AOI and a range of possible numbers of people $i\in \{1, 2, 3, \dots, I\}$. Here, $I$ represents the maximum number of people that are taken into consideration. We generate a total of $C$ cases, where $C = {S \choose 1}+{S \choose 2}+\dots+{S \choose I}$, to cover all the possible scenarios. For each case, we collect CSI data from the predefined $R$ reflected beams of the RIS. This results in the creation of a CSI fingerprinting database with a dimension of $C \cdot R$. The collected CSI data is preprocessed before being employed to train our WiRiS model, which will be explained in the next section.

\subsection{Feature Preprocessing}
After the CSI fingerprinting database is established, the data undergoes preprocessing. To ensure the source data maintains its normal distribution and is not distorted or lost during the calculation, we perform max-min normalization on each CSI amplitude, which is represented as follows:
\begin{equation}
\Bar H_{o, k}^p = \frac{|H_{o, k}^p|-\mathop{\min}\limits_{k}|H_{o, k}^p|}{\mathop{\max}\limits_{k}|H_{o, k}^p|-\mathop{\min}\limits_{k}|H_{o, k}^p|}, \forall k = 1, 2, \cdots, K,
\end{equation}
where $k \in \{1, 2, 3, \dots, K\}$ denotes the subcarrier index, and $p \in \{1, 2, 3, \dots, P\}$ represents the index of each antenna pair. As a result, after switching all predefined reflected beams of the RIS at every location, $R$ unique $\Bar H_{o}^p$ vectors are generated for each antenna pair. Finally, we obtain the input vector of the Transformer model $H_{c}$ by concatenating all the antenna pairs and $R$ different RIS configurations as
\begin{equation}
H_{c} =
\begin{bmatrix}
\Bar H_{o,1,1}^1, \cdots, \Bar H_{o,K,1}^1, \cdots, \Bar H_{o,1,R}^P, \cdots, \Bar H_{o,K,R}^P
\end{bmatrix},
\end{equation}
where $r \in \{1, 2, 3, \dots, R\}$ denotes the reflected beam index of the predefined RIS. Each reflected beam could map to a specific configuration of the RIS.

\subsection{WiRiS System}
In previous indoor localization research, the focus has been primarily on locating a single person in a given space. However, in practical scenarios where multiple people are present, these algorithms are often ineffective. The main challenge in localizing multiple people is determining the exact number of individuals and their respective locations, especially when they are stationary. To address this issue, we introduce the WiRiS system, which is based on the Transformer model \cite{transformer}. The Transformer model offers two key features: a sequence-to-sequence architecture and a multi-head inter-subcarrier attention mechanism, as illustrated in Fig. \ref{fig:block}. We describe how we apply these features to the localization problem for multiple people in the following paragraph.

We begin by explaining the sequence-to-sequence architecture in the Transformer model, which consists of an encoder and a decoder. We consider the RIS-enhanced CSI input of the model as a sentence, with each subcarrier treated as a word. The output includes the embedded position coordinates, with special tokens $<$SOS$>$ and $<$EOS$>$ used to denote the start and end of the sentence output, respectively. The encoder extracts features from the input and converts them into a fixed-length representation. This fixed-length representation, along with the current predicted output, is adopted for predictions. The output length varies based on the number of individuals being localized. For instance, when the input includes CSI from one person, the predicted output should be in a format of ($<$SOS$>$, $<x_{1}>$, $<y_{1}>$, $<$EOS$>$). In contrast, if the input includes CSI features from two individuals, the predicted output will become ($<$SOS$>$, $<x_{1}>$, $<y_{1}>$, $<x_{2}>$, $<y_{2}>$, $<$EOS$>$), and so on. This allows the model to predict the positioning coordinates of different numbers of people simultaneously.

The input dimension of the CSI data grows proportionally with the number of predefined reflected beams, which poses a challenge for predicting the correct number of people and their corresponding locations. To address this challenge, we adopt the multi-head inter-subcarrier attention mechanism in the Transformer model. Within this approach, attention is assigned to subcarriers based on their importance, with more critical subcarriers receiving higher weights. The operation of $i$-th head is represented as
\begin{equation}
head_i = Attention\left( QW_i^Q, KW_i^K, VW_i^V \right)
\end{equation}
where $Q$, $K$, and $V$ are matrices of queries, keys, and values, respectively, with dimensions of $\mathbb{R}^{N_{Q}}$, $\mathbb{R}^{N_{K}}$ and $\mathbb{R}^{N_{V}}$, where $N_{\mathcal{A}}, \mathcal{A}\in\{Q,K,V\}$ indicates the dimension of hidden features. We use matrices $W_i^Q$, $W_i^K$, and $W_i^V$ of dimensions $\mathbb{R}^{N_{Q}\times N_{W_Q}}$, $\mathbb{R}^{N_{K}\times N_{W_K}}$, $\mathbb{R}^{N_{V}\times N_{W_V}}$ respectively to produce various subspace representations of the query, key, and value matrices for all $i = 1, 2, \cdots, N_{enc}$. Note that $N_{W_\mathcal{A}}$ stands for the dimension of hidden features of $W_i^{\mathcal{A}}$, whilst $N_{enc}$ the number of encoding layers of Transformer. The multi-head mechanism enables the attention function to extract information from diverse subspaces, as shown in the following equation:
\begin{equation}
MultiHead(Q, K, V) \!=\! Concat\left(head_1, \cdots, head_{N_{enc}}\right)W^O,
\end{equation}
where $Concat(\cdot)$ means horizontal concatenation operation. Notation of $W^O \in \mathbb R^{N_{O_1} \times N_{O_2}}$ represents the multi-head output with dimension $N_{O_1}$ by $ N_{O_2}$. In the Transformer model, the encoder and decoder comprise multiple sub-layers. We pass the CSI input through four identical encoder sub-layers, each of which has residual connections followed by layer normalization, as shown in the following equation:
\begin{equation}
AddNorm(x) = LayerNorm\Big(x + sublayer(x)\Big),
\end{equation}
where $x$ represents the input of each sub-layer. $LayerNorm(\cdot)$ conducts normalization within a layer and sub-layer. To elaborate a little further, the first part of the decoder's multi-head inter-subcarrier attention is masked to avoid the model from acquiring the entire output label during the training stage.

\begin{figure}
  \centering
  \includegraphics[width=0.47\textwidth]{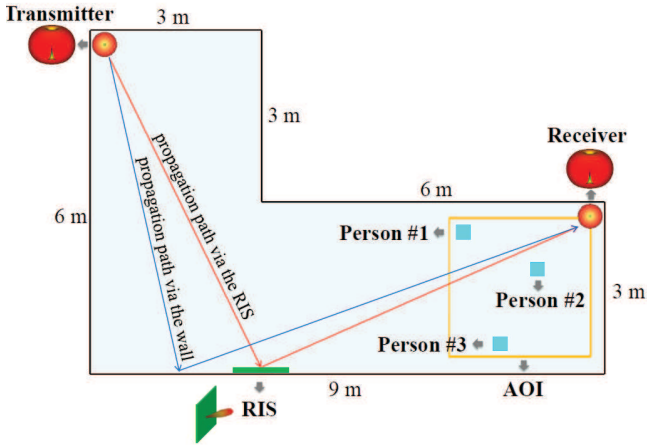}
  \caption{Layout of the RIS ray-tracing scenario with multi-people in AOIs. Blue ray propagates over a wall, whereas red ray impinges RIS reflecting signals to the receiver.}
  \label{fig:block2}
\end{figure}

\begin{figure}
  \centering
  \begin{subfigure}[b]{0.46\textwidth}
    \includegraphics[width=\textwidth]{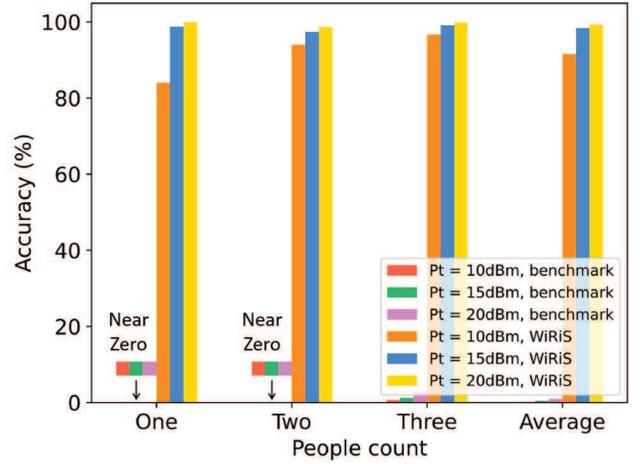}
    \caption{\footnotesize}
    \label{fig:sim1}
  \end{subfigure}
  \begin{subfigure}[b]{0.46\textwidth}
    \includegraphics[width=\textwidth]{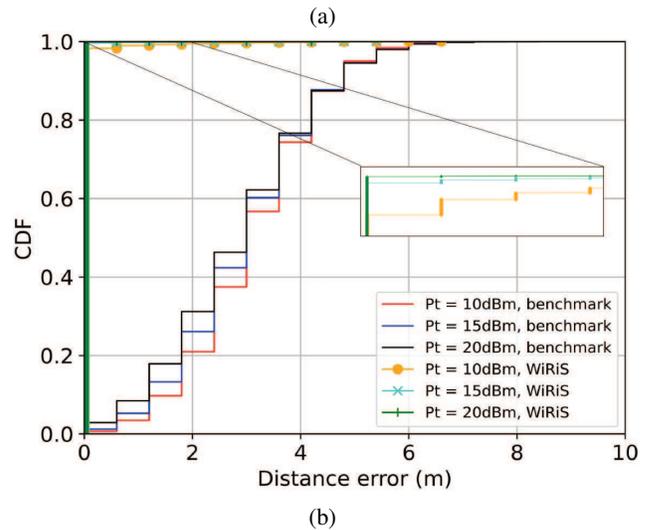}
    \caption{\footnotesize}
    \label{fig:sim2}
  \end{subfigure}
\caption{Simulation results of WiRiS: (a) Accuracy of people counting, and (b) CDF of positioning distance error.}
\label{fig:simulation1}
\end{figure}

\section{Performance Evaluation}

\subsection{Simulation Setting and Results}
In our simulation, we assume that the LoS path between the transmitter and receiver is blocked. The operating frequency of Wi-Fi is at $5.6$ GHz as well as RIS. To improve signal coverage in the AOI and enhance the human presence effect, we deploy a feasible size of $1 \times 1 m^{2}$ RIS posted on the wall in an L-shaped room, as illustrated in Fig. \ref{fig:block2}. The reflected beams of the RIS are mapped to $9$ reflection angles ranging from $0^\circ$ to $80^\circ$ with each having $10^\circ$ half-power beamwidth. We position the transmitter and receiver in different corners of the space, each equipped with a single omnidirectional antenna. We set the Transformer parameters in WiRiS as $N_{Q} = N_{K} = N_{V}=512$, $N_{W_Q} = N_{W_K} = N_{W_V}=64$, and $N_{O_1}=N_{O_2}=512$ to facilitate the concatenation operation. We use $N_{enc}=8$ encoding layers and $N_{dec}=4$ decoding layers. To generate the CSI data, we use the 3D ray-tracing software Wireless InSite \cite{WI} to simulate the scenario and generate $16$ reference points uniformly distributed in the AOI, with 1 to 3 people present in each region. We generate total ${16 \choose 1} + {16 \choose 2} + {16 \choose 3} = 696$ cases of CSI data labels for the scenario without RIS. With the RIS's $9$ predefined reflected beams, there will be a total of $6264$ cases, which is computed from $696$ times the number of reflected RIS beams. The whole data generation and training are performed under GeForce RTX 4090 graphics processing unit (GPU). It is worth mentioned that we only compare the proposed WiRiS with benchmark without deployment of RIS. Due to no papers focused on joint sensing and localization, we have no other benchmarks investigated.

We evaluate the performance of the WiRiS system using simulation data, as depicted in Fig. \ref{fig:simulation1}. Fig. \ref{fig:sim1} shows the people counting results in scenarios with and without the RIS, with transmit power levels of $P_t=\{10,15,20\}$ dBm. In the absence of the RIS, the people counting accuracy is near zero in arbitrary cases. In contrast, when employing the RIS, the people counting accuracy exceeds $90\%$ for all cases. We can infer that under NLoS accurately predicting the number of people without RIS is difficult. Fig. \ref{fig:sim2} compares the cumulative distribution function (CDF) for positioning error when the people counting result is correct. The results demonstrate that all three cases with the RIS exhibit a comparably lower positioning error than the other three cases without the RIS. In summary, our findings indicate that people counting accuracy and positioning error are proportional to transmit power. The presence of the RIS significantly improves the people counting accuracy and reduces the positioning error, especially in scenarios with larger pathloss.

\subsection{Experiment Setting and Results}

\begin{figure}
  \centering
  \begin{subfigure}[c]{0.46\textwidth}
    \includegraphics[width=\textwidth]{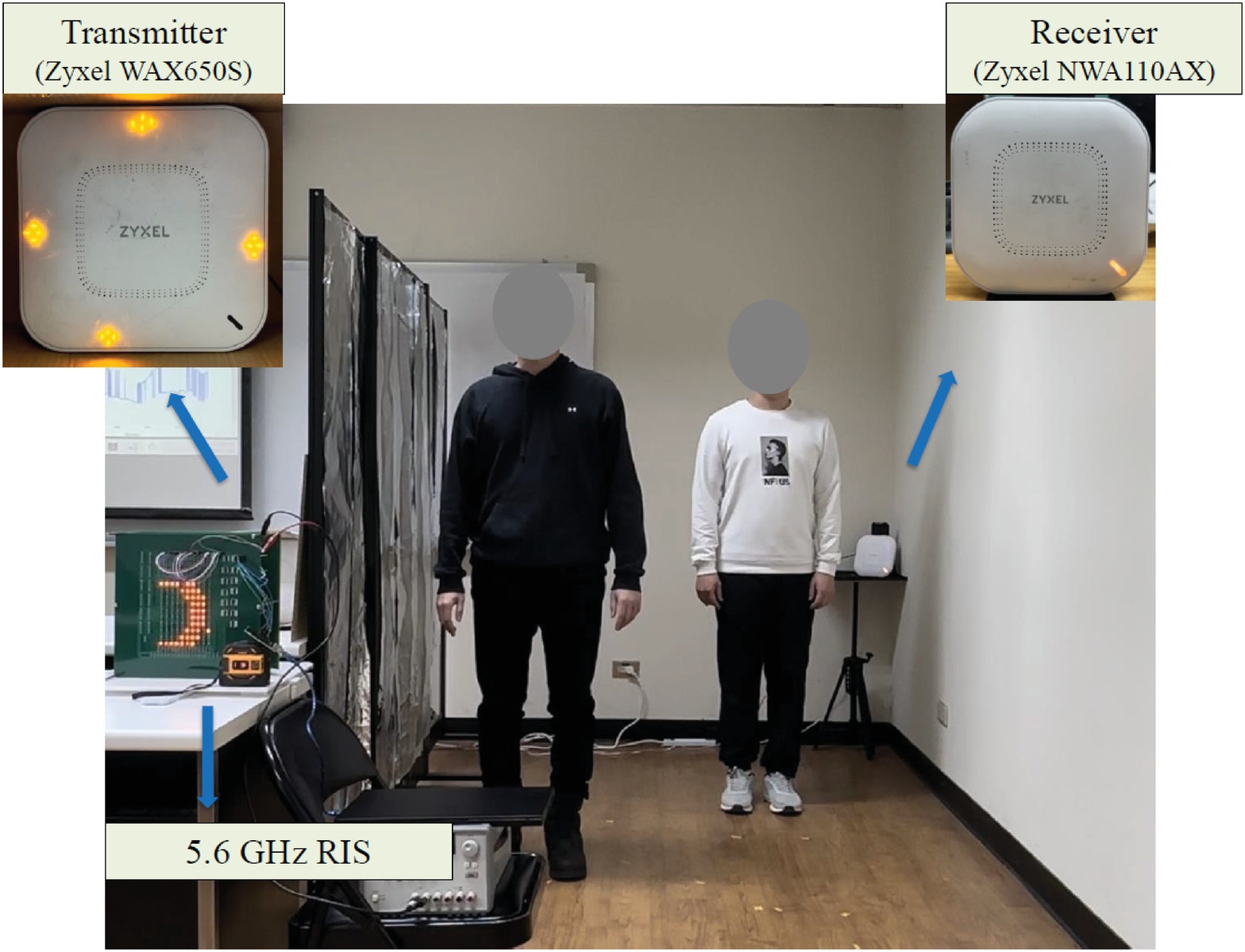}
    \caption{\footnotesize}
    \label{fig:layout}
  \end{subfigure}
  \begin{subfigure}[c]{0.23\textwidth}
    \includegraphics[width=\textwidth]{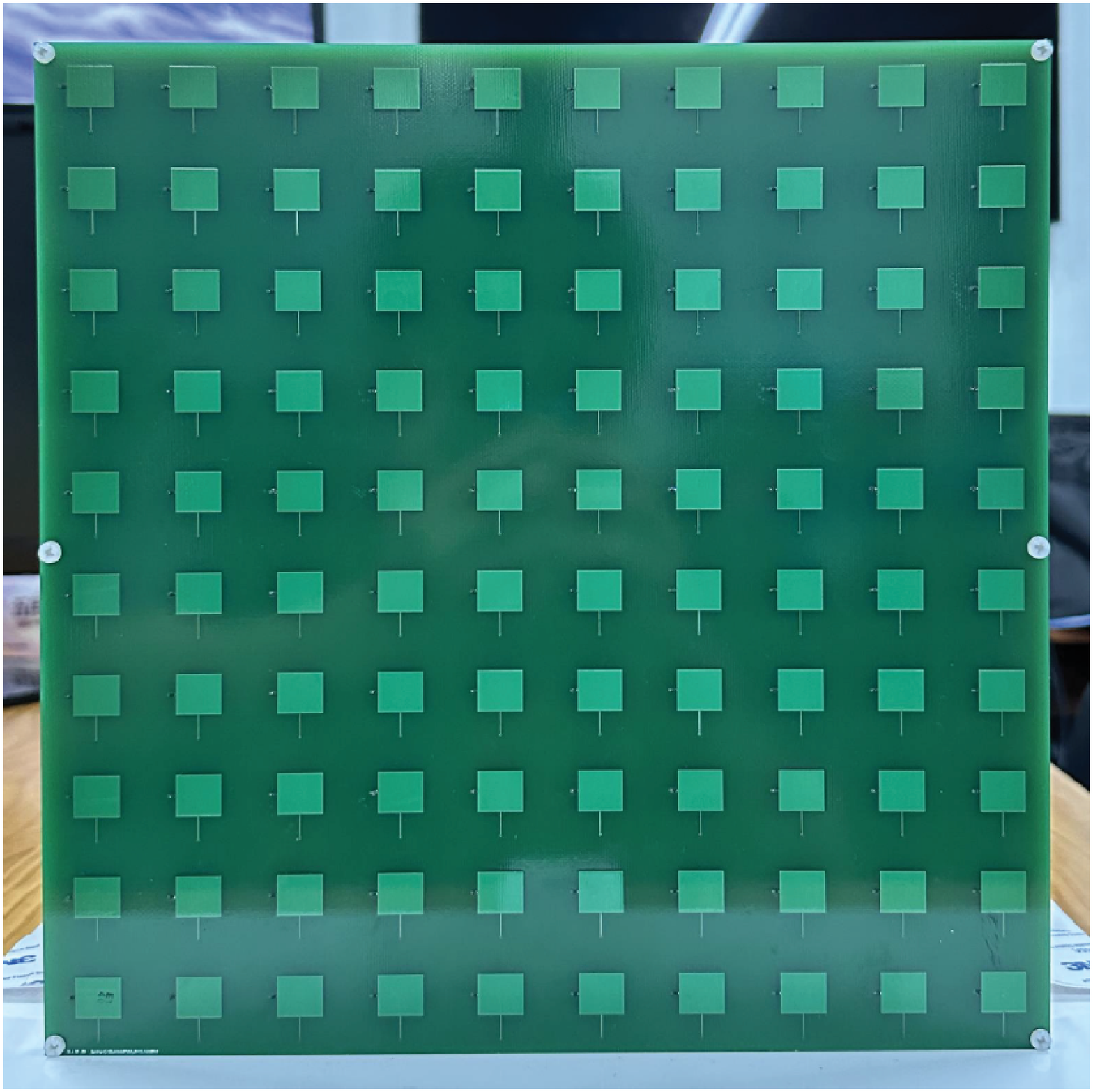}
    \caption{\footnotesize}
    \label{fig:risf}
  \end{subfigure}
  \begin{subfigure}[c]{0.23\textwidth}
    \includegraphics[width=\textwidth]{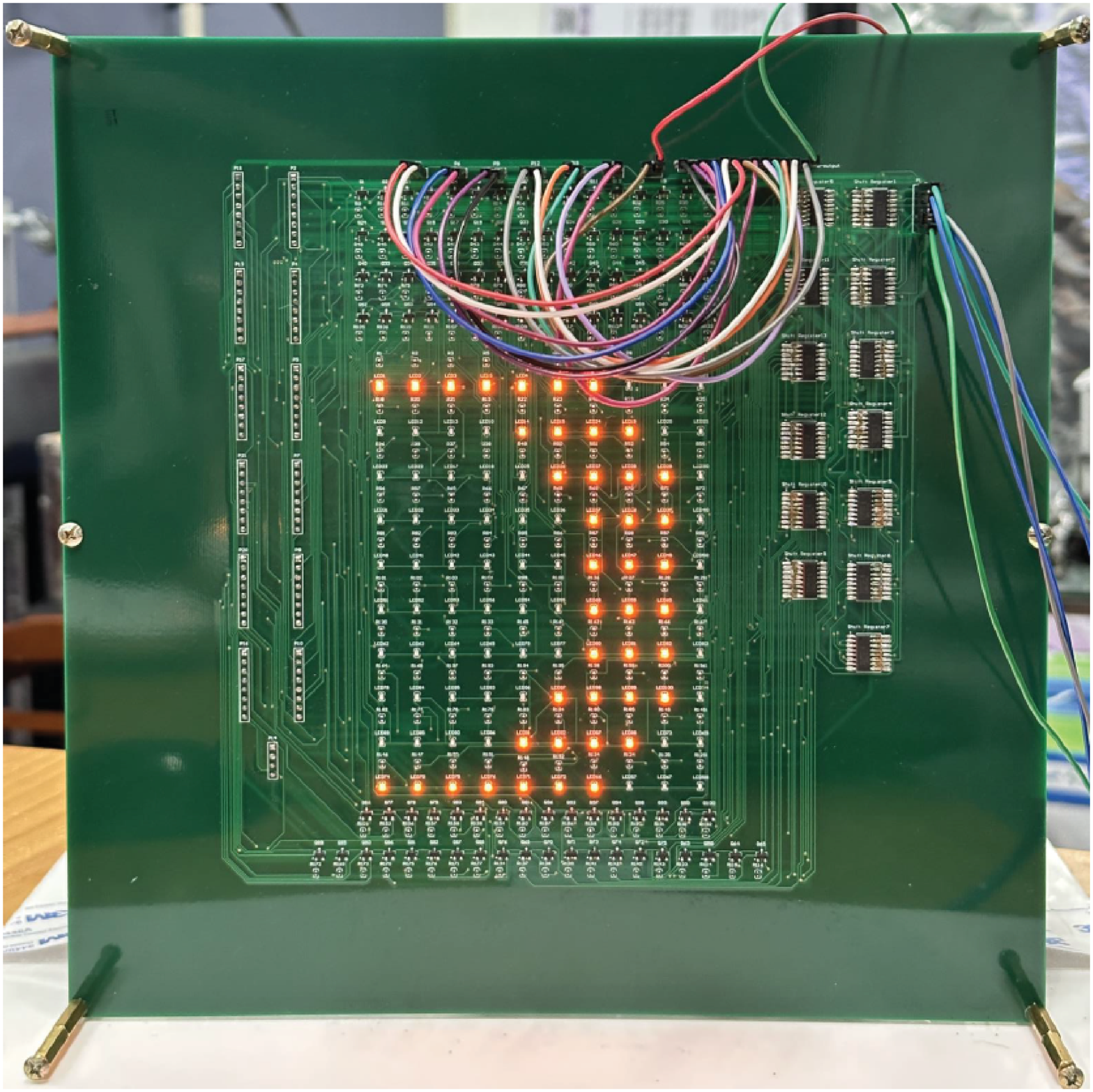}
    \caption{\footnotesize}
    \label{fig:risb}
  \end{subfigure}
\caption{Experimental setup of WiRiS: (a) Test field with 5.6 GHz Wi-Fi transmitter/receiver and RIS, (b) frontside view of RIS, and (c) backside view of RIS.}
\label{fig:experiment}
\end{figure}

As illustrated in Fig. \ref{fig:experiment}, we conducted our experiment in a $2 \times 2.5 m^{2}$ area within a conference room layout in Fig. \ref{fig:layout}. We have adopted a Wi-Fi based single-antenna transmitter ZYXEL WAX650S \cite{WAX650S} and a dual-antenna receiver ZYXEL NWA110AX \cite{NWA110AX} operating at $5.6$ GHz with a $20$ MHz bandwidth. The transmit power is set as $10$ dBm. A prototype of single RIS \cite{ccu} is placed $1m$ in front of the transmitter, comprising $100$ elements, each with $1$-bit of two states (ON/OFF), as depicted in Fig. \ref{fig:risf}. Note that the codebooks of selected beamforming patterns are offline configured via bio-inspired heuristic schemes. We set the incidence angle ${\theta_i}$ to $0^\circ$ and select the reflection angles of ${\theta_r} = \{10^\circ, 20^\circ, 30^\circ, 40^\circ, 50^\circ, 60^\circ\}$ for the RIS. The RIS controller displays the corresponding element state on an LED when it switches to the designated reflection angle, i.e., LED lights up for ON and vice versa for OFF as shown in Fig. \ref{fig:risb}. To simulate a blockage scenario, we place a whiteboard with aluminium foil between the transmitter and receiver. We select four reference points with a spacing of $1m$ during the offline phase, and consider the presence of ${1,2,3}$ people in the AOI without portable devices. Such setting provides a total of ${4 \choose 1}+{4 \choose 2}+{4 \choose 3} = 14$ candidate labels. We collect CSI data in two scenarios to compare the performance with and without the RIS. In the scenario without the RIS, there were $14$ cases, while in the scenario with the RIS, we switch the RIS with six beams toward the AOI resulting in a total of $14 \times 6 = 84$ cases. We have collected $1000$ CSI data per case as the database in both scenarios, and split the datasets into $600$ for training, $200$ for validation and $200$ for testing data.

\begin{figure}
  \centering
  \begin{subfigure}[c]{0.23\textwidth}
    \includegraphics[width=\textwidth]{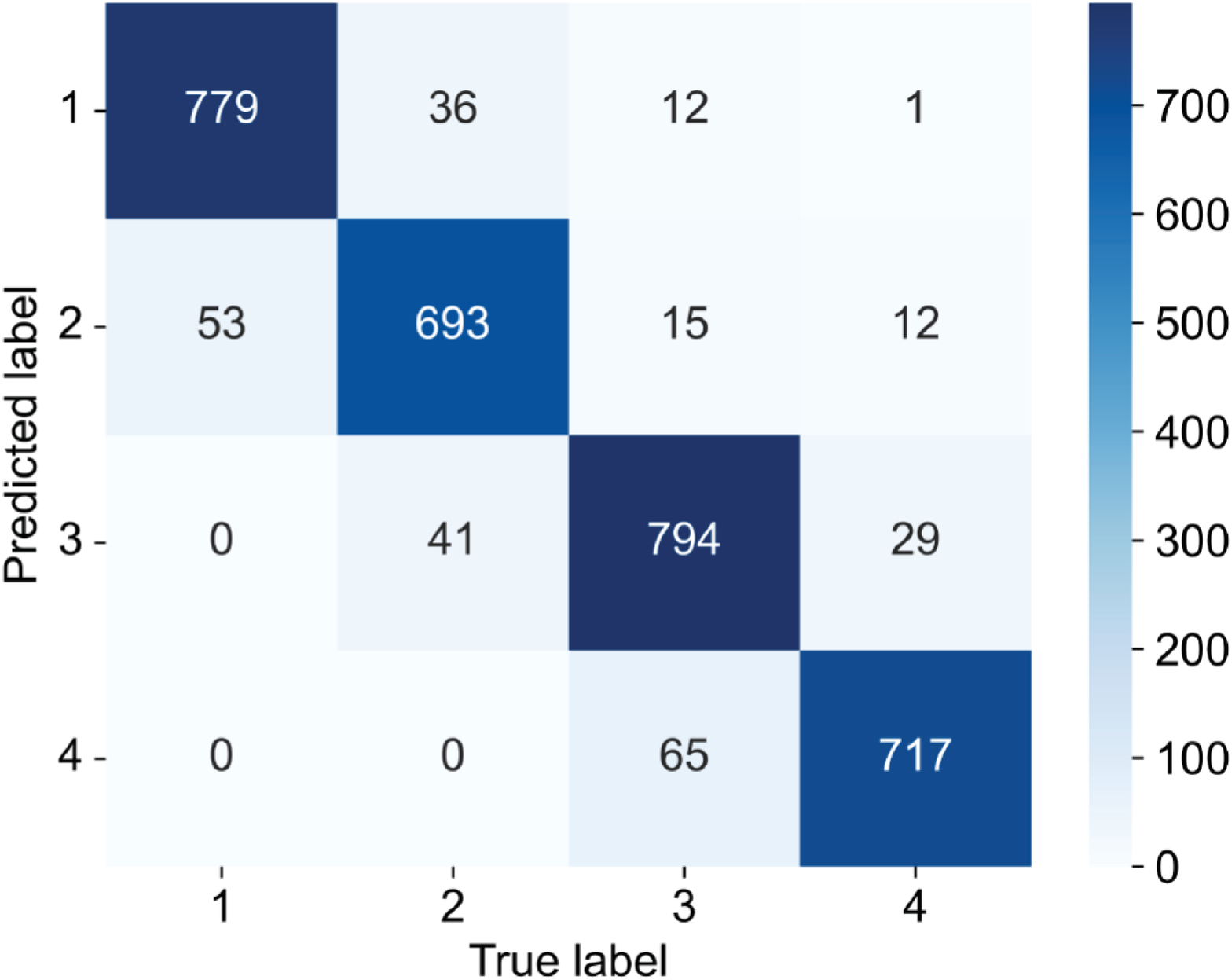}
    \caption{\footnotesize}
    \label{fig:exp3}
  \end{subfigure}
  \quad
  \begin{subfigure}[c]{0.23\textwidth}
    \includegraphics[width=\textwidth]{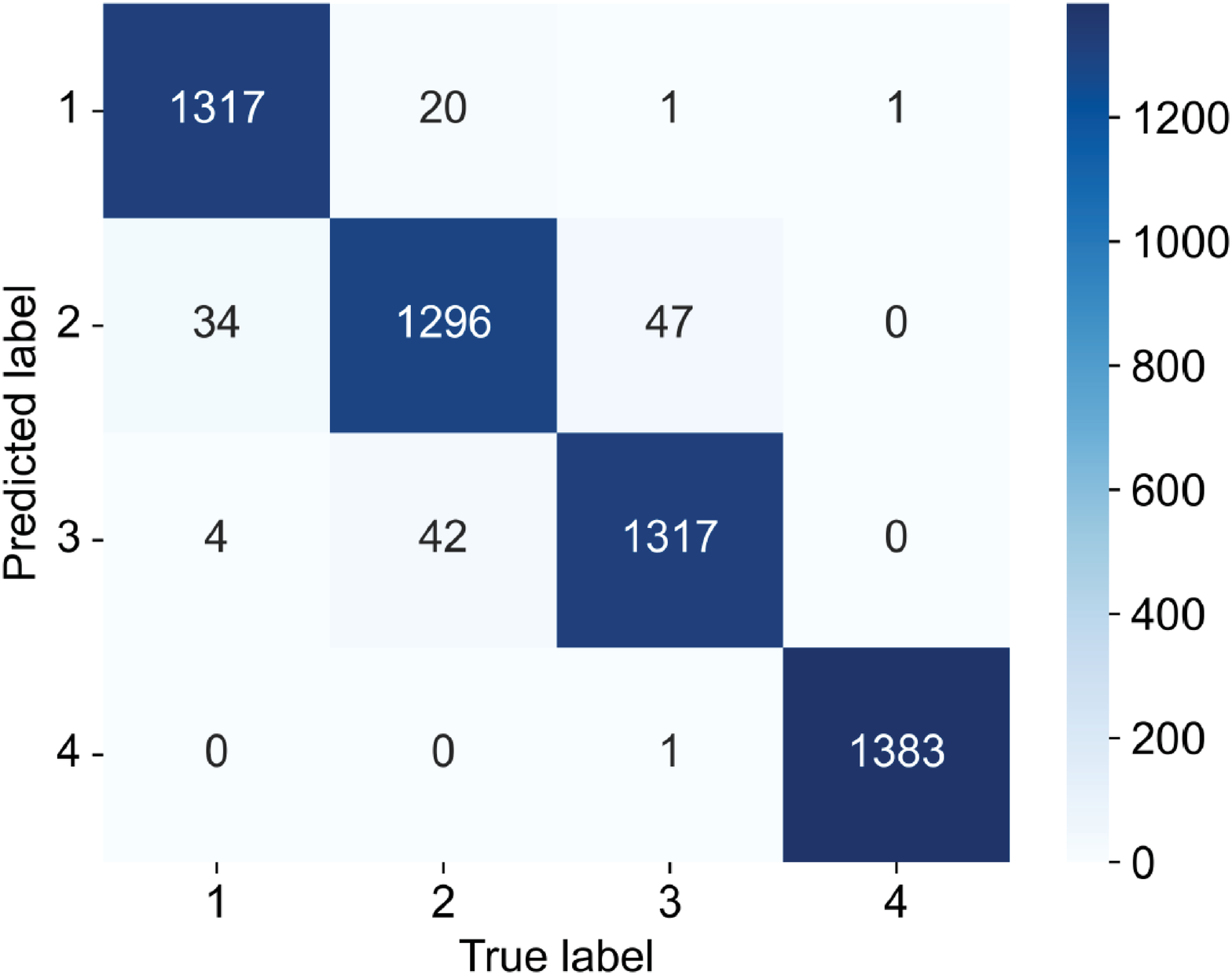}
    \caption{\footnotesize}
    \label{fig:exp4}
  \end{subfigure}
\caption{Confusion matrices of (a) benchmark and (b) proposed WiRiS system.}
\label{fig:experiment2}
\end{figure}

\begin{figure}
  \centering
    \includegraphics[width=0.48\textwidth]{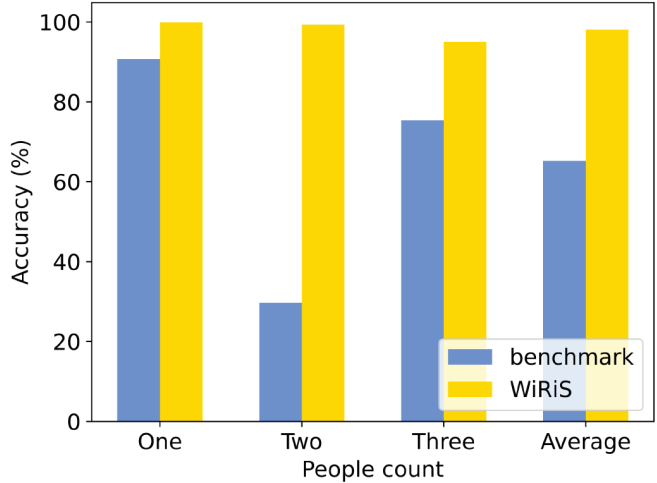}
\caption{Experiment results of proposed WiRiS system in terms of accuracy of people counting.}
\label{fig:experiment1}
\end{figure}

Fig. \ref{fig:experiment2} illustrates the confusion matrix of localization classification under the correct people count with benchmark without RIS in Fig. \ref{fig:exp3} and WiRiS system in Fig. \ref{fig:exp4}. The index of the true label is the ground truth of the predefined reference points. According to the results, the performance is better in term of people counting and location classification with the assistance of RIS. In the following, we evaluate the proposed WiRiS system by the collected data from the experiment. As shown in Fig. \ref{fig:experiment1}, with the assistance of RIS, the accuracy increased from $90.75\%$ to $99.88\%$ with one person, increased from $29.67\%$ to $99.33\%$ with two people, and increased from $75.38\%$ to $95\%$ with three people. In the benchmark scenario without RIS, since there are more cases with two people, the accuracy in distinguishing between two people is lower. Comparing the results of accuracy of WiRiS and benchmark without the RIS, the average people counting accuracy improved from $65.26\%$ to $98.07\%$.

\section{Conclusion}
In this paper, we have presented a novel approach WiRiS for joint people counting and localization in the indoor NLoS region. Our approach leverages a predefined RIS to direct the signal towards the NLoS area, thereby improving the localization accuracy. Transformer-based learning technique is employed to simultaneously provide discrete and continuous determinations. We have collected CSI data during the offline phase, which helps to effectively separate the location features between neighboring reference points. We also consider the presence of multiple people uniformly distributed on the reference points. During the online phase, the number of people and their locations are estimated by switching all reflected beams of the RIS. Results from both simulations and prototype experiments show that the proposed WiRiS system achieves over $90\%$ accuracy in people counting and localization with the assistance of RIS, which outperforms the existing benchmark without the RIS deployment.

\bibliographystyle{IEEEtran}
\bibliography{reference}

\begin{thebibliography}{10}
\providecommand{\url}[1]{#1}
\csname url@samestyle\endcsname
\providecommand{\newblock}{\relax}
\providecommand{\bibinfo}[2]{#2}
\providecommand{\BIBentrySTDinterwordspacing}{\spaceskip=0pt\relax}
\providecommand{\BIBentryALTinterwordstretchfactor}{4}
\providecommand{\BIBentryALTinterwordspacing}{\spaceskip=\fontdimen2\font plus
\BIBentryALTinterwordstretchfactor\fontdimen3\font minus
  \fontdimen4\font\relax}
\providecommand{\BIBforeignlanguage}[2]{{%
\expandafter\ifx\csname l@#1\endcsname\relax
\typeout{** WARNING: IEEEtran.bst: No hyphenation pattern has been}%
\typeout{** loaded for the language `#1'. Using the pattern for}%
\typeout{** the default language instead.}%
\else
\language=\csname l@#1\endcsname
\fi
#2}}
\providecommand{\BIBdecl}{\relax}
\BIBdecl

\bibitem{Low_efforts}
X.~Tong, Y.~Wan, Q.~Li, X.~Tian, and X.~Wang, ``{CSI} fingerprinting
  localization with low human efforts,'' \emph{IEEE/ACM Transactions on
  Networking}, vol.~29, no.~1, pp. 372--385, 2021.

\bibitem{IoT}
R.~Y. Chang, S.~J. Liu, and Y.~K. Cheng, ``Device-free indoor localization
  using {Wi-Fi} channel state information for internet of things,'' \emph{in
  Proceedings IEEE Global Communications Conference}, pp. 1--7, 2018.

\bibitem{Radio_img}
Q.~Gao, J.~Wang, X.~Ma, X.~Feng, and H.~Wang, ``{CSI}-based device-free
  wireless localization and activity recognition using radio image features,''
  \emph{IEEE Transactions on Vehicular Technology}, vol.~66, no.~11, pp.
  10\,346--10\,356, 2017.

\bibitem{ML_RSSI}
N.~Singh, S.~Choe, and R.~Punmiya, ``Machine learning based indoor localization
  using {Wi-Fi RSSI} fingerprints: An overview,'' \emph{IEEE Access}, vol.~9,
  pp. 127\,150--127\,174, 2021.

\bibitem{CSI_Loc}
X.~Wang, L.~Gao, S.~Mao, and S.~Pandey, ``{CSI}-based fingerprinting for indoor
  localization: A deep learning approach,'' \emph{IEEE Transactions on
  Vehicular Technology}, vol.~66, no.~1, pp. 763--776, 2017.

\bibitem{NLOS_Color}
C.-C. Hsieh, A.-H. Hsiao, C.-J. Chiu, and K.-T. Feng, ``{CSI} ratio with
  coloring-assisted learning for {NLoS} motionless human presence detection,''
  \emph{in Proceedings IEEE Vehicular Technology Conference}, pp. 1--5, 2022.

\bibitem{Improve_NLOS}
K.~Bregar and M.~Mohorcic, ``Improving indoor localization using convolutional
  neural networks on computationally restricted devices,'' \emph{IEEE Access},
  vol.~6, pp. 17\,429--17\,441, 2018.

\bibitem{my}
\BIBentryALTinterwordspacing
L.-H. Shen, C.-C. Hsieh, A.-H. Hsiao, and K.-T. Feng, ``{CRONOS}: Colorization
  and contrastive learning enhanced {NLoS} human presence detection using
  {Wi-Fi CSI} signals,'' \emph{arXiv e-Print}, 2022. [Online]. Available:
  \url{https://arxiv.org/abs/2211.10354}
\BIBentrySTDinterwordspacing

\bibitem{acm}
L.-H. Shen, K.-T. Feng, and L.~Hanzo, ``Five facets of {6G}: Research
  challenges and opportunities,'' \emph{ACM Computing Surveys}, vol.~55,
  no.~11, pp. 1--39, 2023.

\bibitem{my2}
C.-J. Ku, L.-H. Shen, and K.-T. Feng, ``Reconfigurable intelligent surface
  assisted interference mitigation for {6G} full-duplex {MIMO} communication
  systems,'' in \emph{Proceedings IEEE Personal, Indoor and Mobile Radio
  Communications (PIMRC)}, 2022, pp. 327--332.

\bibitem{WFing}
C.~L. Nguyen, O.~Georgiou, G.~Gradoni, and M.~D. Renzo, ``Wireless
  fingerprinting localization in smart environments using reconfigurable
  intelligent surfaces,'' \emph{IEEE Access}, vol.~9, pp. 135\,526--135\,541,
  2021.

\bibitem{MetaRadar}
H.~Zhang, J.~Hu, H.~Zhang, B.~Di, K.~Bian, Z.~Han, and L.~Song, ``{MetaRadar}:
  Indoor localization by reconfigurable metamaterials,'' \emph{IEEE
  Transactions on Mobile Computing}, vol.~21, no.~8, pp. 2895--2908, 2022.

\bibitem{SEMI}
K.~Keykhosravi, M.~F. Keskin, S.~Dwivedi, G.~Seco-Granados, and H.~Wymeersch,
  ``Semi-passive 3d positioning of multiple {RIS}-enabled users,'' \emph{IEEE
  Transactions on Vehicular Technology}, vol.~70, no.~10, pp. 11\,073--11\,077,
  2021.

\bibitem{Towards}
H.~Zhang, H.~Zhang, B.~Di, K.~Bian, Z.~Han, and L.~Song, ``Towards ubiquitous
  positioning by leveraging reconfigurable intelligent surface,'' \emph{IEEE
  Communications Letters}, vol.~25, no.~1, pp. 284--288, 2021.

\bibitem{POE}
A.~Elzanaty, A.~Guerra, F.~Guidi, and M.~S. Alouini, ``Reconfigurable
  intelligent surfaces for localization: Position and orientation error
  bounds,'' \emph{IEEE Transactions on Signal Processing}, vol.~69, pp.
  5386--5402, 2021.

\bibitem{RF_sensing}
J.~Hu, H.~Zhang, B.~Di, L.~Li, K.~Bian, L.~Song, Y.~Li, Z.~Han, and H.~V. Poor,
  ``Reconfigurable intelligent surface based {RF} sensing: Design,
  optimization, and implementation,'' \emph{IEEE Journal on Selected Areas in
  Communications}, vol.~38, no.~11, pp. 2700--2716, 2020.

\bibitem{Meta_sensing}
J.~Hu, H.~Zhang, K.~Bian, M.~D. Renzo, Z.~Han, and L.~Song, ``Metasensing:
  Intelligent metasurface assisted {RF} 3d sensing by deep reinforcement
  learning,'' \emph{IEEE Journal on Selected Areas in Communications}, vol.~39,
  no.~7, pp. 2182--2197, 2021.

\bibitem{transformer}
A.~Vaswani, N.~Shazeer, N.~Parmar, J.~Uszkoreit, L.~Jones, A.~N. Gomez,
  {\L}.~Kaiser, and I.~Polosukhin, ``Attention is all you need,'' in
  \emph{Proceedings ACM Advances in Neural Information Processing Systems},
  2017, pp. 5998--6008.

\bibitem{WI}
\BIBentryALTinterwordspacing
{Remcom, Wireless Insite}. [Online]. Available:
  \url{https://www.remcom.com/wireless-insite-em-propagation-software/}
\BIBentrySTDinterwordspacing

\bibitem{WAX650S}
\BIBentryALTinterwordspacing
{Zyxel, WAX650S}. [Online]. Available:
  \url{https://www.zyxel.com/tw/zh/products/wireless/80211ax-wifi-6-dual-radio-unified-pro-access-point-wax650s/specification}
\BIBentrySTDinterwordspacing

\bibitem{NWA110AX}
\BIBentryALTinterwordspacing
{Zyxel, NWA110AX}. [Online]. Available:
  \url{https://www.zyxel.com/tw/zh/products/wireless/80211ax-wifi-6-dual-radio-poe-access-point-nwa110ax/specification}
\BIBentrySTDinterwordspacing

\bibitem{ccu}
Y.-C. Lin, C.-T. Lin, Y.-C. Lin, W.-Y. Chen, S.-F. Chang, and S.-C. Lin,
  ``Effects of large angle of incidence in offset-fed reflectarray antennas,''
  in \emph{Proceedings IEEE International Symposium on Antennas and Propagation
  and USNC-URSI Radio Science Meeting}, 2019, pp. 297--298.

\end{thebibliography}

\end{document}